\documentstyle[aps,epsfig,color,preprint,amssymb,subfigure]{revtex}
\tighten 
\newcommand{\bcols}{\ifpreprintsty\else\begin{multicols}{2}\fi}
\newcommand{\ecols}{\ifpreprintsty\else\end{multicols}\fi}

\begin{document}
\draft

\title {Molecular wire-nanotube interfacial effects on electron transport} 
\author {Giorgos Fagas, Gianaurelio Cuniberti,
and Klaus Richter}
\address {Max-Planck-Institut f{\"u}r Physik komplexer Systeme,
N{\"o}thnitzer Strasse 38, D-01187 Dresden, Germany}
\date{\today}
\maketitle
\begin{abstract}
We discuss the conductance of a molecular bridge between
mesoscopic electrodes supporting low-dimensional transport
and bearing an internal structure. As an example for such nanoelectrodes
we assume semi-infinite~(carbon) nanotubes.  In the Landauer scattering matrix approach,
we show that the conductance of this hybrid is very sensitive to the geometry
of the contact unlike the usual behaviour in the presence of bulk electrodes.

\end{abstract}
\bcols

The urge for smaller, faster, and cheaper electronic devices drives a race
of increasing pace for the miniaturisation of transistors. Such an
activity has provided an interplay between fundamental and applied research
giving rise to the fields of mesoscopic and nano physics. The principal ideas are based on
the `top-down' miniaturisation of electronic devices to micro- or
nanometre-size and the exploitation of nontrivial quantum effects in
solid state physics at these length scales. An arena closely related to
mesoscopics is molecular electronics~\cite{AR98}.
As in mesoscopic systems, quantum mechanical methods need to be accommodated as
compared to the conventional description of electronic devices.
In addition, a detailed microscopic analysis of the physical objects
involved should be included since part of the challenge is to use
individual molecules or supramolecular structures as (reproducible) circuit
elements. This is the `bottom-up' manufacture philosophy.

Since the original proposal of a molecular rectifier \cite{AR74} significant progress
on transport across a single molecule has been demonstrated experimentally only in recent years.
This owes to the advances in self-assembly techniques, end-group modifications,
scanning probe and break-junction techniques, which allow
atomic-scale control and positioning of single molecules and their assemblies.
As a result, a series of electron transport measurements through molecular
complexes between metallic pads has been reported~\cite{RZMBT97,FS99}.

Such experiments complemented by theoretical efforts have pointed out a number of
factors that determine the conductance in such
structures~\cite{RZMBT97,FS99,MKR94a,STDHK96,JV96,EK98a,YR98}.
The most important include the
molecular electronic structure related to the resonant spectrum and the
distribution of the wavefunction along the bridge, the location of the
electrodes equilibrium Fermi energy with respect to the molecular electronic spectrum,
charging effects, the electron-phonon coupling in wires of low conduction,
and the interface between electrodes and molecular complexes.
Many of these effects are inter-related but the classification helps in
understanding the basic underlying transport mechanisms.

In this paper, we focus on the effect of the molecular wire-electrode interface on electron transport.
In a typical experimental setup a molecular bridge connects two electrodes
acting as electron donor and acceptor reservoirs. Despite this conceptual separation of the
molecular device into its constituents  what is measured
is the conductance across the whole system. This implies that
besides the intrinsic molecular ability to convey charge
the coupling of the molecular complex to the environment is significant.
This argument is supported by theoretical work which shows the need for an
atomistic description~\cite{EK98a,YR98}.

Moreover, in theoretical studies one usually assumes the molecular wire
attached to one dimensional or bulky electrodes~\cite{MKR94a,STDHK96,JV96,EK98a,YR98}.
Both are described by a smooth local density of states, whereas, the latter bear also a continuum
of possible conducting channels. But none of these conditions may hold in general.
For example such a description is inadequate for experiments employing
scanning tunnelling microscope tips where interface-induced effects
have been found to strongly influence the measured conductance~\cite{STMG}.
The tip electronic structure, which is determined by either structural
formation or adsorbates, gives rise to characteristic signatures. To account for such effects
one may study the conductance across a cluster which includes the molecule and the tip~\cite{EK98a}.
In contrast to previous studies we assume carbon nanotubes which
exemplify the contrary of both aforementioned features. Apart from being an
ideal example, carbon tubules are also promoted as electrodes with lateral
dimensions comparable to the size of the molecular connector.
Our final aim is to both point out the basic issues and understand how to manipulate the
contact in order to achieve control over electron transport.

Carbon nanotubes support up to two channels for electrons with energy around the
equilibrium Fermi energy and have a complex topology. They are
known to exhibit a wealth of properties depending on
their diameter~($\sim nm$), chirality~(orientation of graphite sheet roll up), and
whether they consist of a single cylindrical surface~(single-wall)
or more~(multi-wall)~\cite{SDD98}.
Carbon tubules are defined by a pair of integers $(m,l)$ denoting the chiral vector.
First experiments to build nanotube-supramolecule-nanotube
hybrids have been attempted~\cite{bachtoldprivate}. Furthermore,
carbon nanotubes are already utilised as scanning probe tips to study
molecular structures~\cite{WJWCL98}.

In what follows, we show that electron transport shares distinct properties
depending on the number of atomic contacts at the interface between the molecular bridge and the
electrodes. It also depends on the symmetry of the channel wavefunctions transverse
to the transport direction.
We demonstrate that single-atom contacts give rise to complex conductance
spectra exhibiting quantum features of both the molecule and the electrodes.
These are attributed to the electronic structure of the molecular wire and to
the local density of states of the leads, respectively. Multiple contacts
provide a mechanism for transport channel selection, leading to a scaling law
for the conductance and allowing for its control.
Channel selection also highlights the role of molecular resonant states
by suppressing details assigned to the electrodes.

We first give specific details of the system we study. The electrodes are open-ended single-wall
carbon nanotubes taken as either armchair or zigzag type. The latter are defined
by the  $(m,m)$ and $(m,0)$ chiral vectors, respectively. For the description of
the electronic spectrum we use a parametrised tight-binding Hamiltonian with a
$\pi$-electron per atom~\cite{SDD98}. This ignores curvature induced effects but takes into
account all qualitative features. Reflecting our aim to provide a qualitative understanding 
of the phenomenology of the physical problem, the molecular wire is also modelled at the
tight-binding level by taking a dimerised chain of $2N$ atomic sites~\cite{JV96}. This is
represented by an alternating nearest-neighbour interaction between subsequent pair of
atoms. The ratio of the coupling strengths determines the dimerisation parameter.
Such a system bears two electronic bands, whereas the width of the bands and the corresponding
gap are determined by the hopping terms.

\begin{figure}[t]
\begin{center}
\subfigure{\epsfig{file=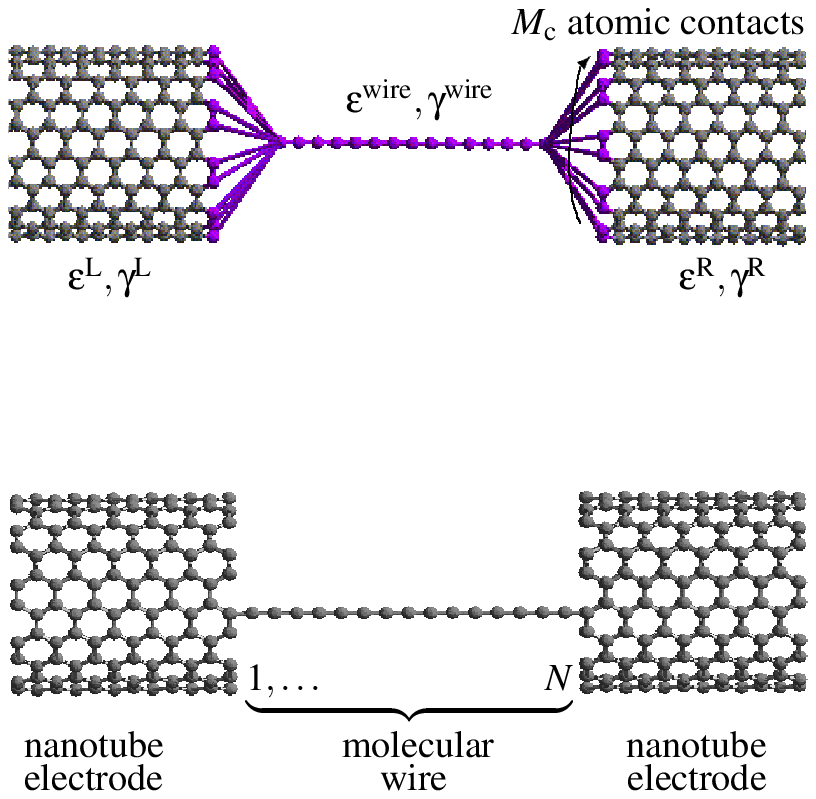, width=.99\linewidth}}
\end{center}

\caption{\label{fig0}
	Scheme of the molecular wire-carbon nanotube hybrid with
	single (bottom) and multiple (top) contacts. In this paper, on-site energies
	$\epsilon^{\alpha={\rm L,R,wire}}$ are fixed to zero.	}
\end{figure}

The electronic Hamiltonian of the system, including the left~(L) and
right~(R) tube (see Fig.~\ref{fig0}), reads

\begin{eqnarray}
\label{meq1}
 H &=& H_{\rm tubes} + H_{\rm wire} + H_{\rm coupling}
\\ \nonumber &=&
-\sum_{\alpha={\rm L,R}} \; \sum_{n^{\phantom{\prime}}_{\alpha},n_{\alpha}^\prime} 
 \gamma^\alpha_{\left\langle n^{\phantom{\prime}}_{\alpha},n_{\alpha}^\prime \right\rangle}
\left| n^{\phantom{\prime}}_{\alpha} \right\rangle 
\left\langle n_{\alpha}^\prime \right| 
\\& - & \sum_{n^{\phantom{\prime}}_{wire}=\;odd}
\gamma^{wire}
\left|n^{\phantom{\prime}}_{wire}\right\rangle 
\left\langle  n_{wire} +1\right|
-\sum_{n^{\phantom{\prime}}_{wire}=\;even}
2\;\gamma^{wire}
\left|n^{\phantom{\prime}}_{wire}\right\rangle 
\left\langle n_{wire}  +1 \right|
\\ & - & \sum_{m_{\rm L}} \Gamma \left| m_{\rm L}\right. \left.\!\! \rangle 
 \langle n_{\rm wire}\!=\! 1\right.\left.\!\! \right|
 \nonumber
 - \sum_{m_{\rm R}} \Gamma \left| m_{\rm R}\right. \left. \! \! \rangle 
 \langle n_{\rm wire}\!=\! N\right.\left.\!\! \right|
+ {\rm H.c.} . \nonumber
\end{eqnarray}

Here, $\gamma^{\rm L,R}$($=2.66$ eV), $\gamma^{\rm wire}$($=1 eV$), and $\Gamma$ denote
nearest-neighbour hopping terms between atoms of the left or right carbon
tubule leads, molecular bridge, and the bridge/lead interface, respectively.
We shall leave $\Gamma$ unspecified for it does not influence the qualitative
behaviour we point out below.
Effects related to the exact magnitude of the interfacial coupling strength
have been described elsewhere \cite{PRB}.
Note that the dimerisation parameter of the wire which is defined as the absolute value of the
difference between the alternating bonds over the average coupling
strength has been fixed to around $0.67$. Our parametrisation sets the band gap and each
bandwidth of the molecular wire to 2 and 3 eV, respectively.
For simplicity, all on-site energies
have been fixed to zero implying that the symmetric band structures of both the carbon
nanotubes and the dimerised chain are centred at zero.
In Eq.~(\ref{meq1}), $n_{\rm L,R}$ is a two-dimensional coordinate
spanning the tube lattice.
Summations over $m_{\rm L}$ and $m_{\rm R}$ run over interfacial end-atoms of the leads.
In general, there are $M$ such atomic positions, depending on the perimeter
of the tubes, and a number of hybridisation contacts ranging from
a single contact $M_{\rm c}=1$ (SC), to multiple contacts
$M_{\rm c} = M$ (MC). 

We also use a square lattice model of mesoscopic electrodes with nearest-neighbour
interactions~($\gamma^{\rm L,R}\!=\!1$eV) and periodic boun\-dary conditions
on the longitudinal cuts parallel to the lattice bonds. Although this model may be
unphysical or at least very difficult to realise experimentally~\cite{SE01}, it is
instructive upon comparison and delivers additional analytic
insight~\cite{JCP}.

For solving the transport problem we use the Landauer theory~\cite{IL99}
which relates the conductance of a system to the electron
transmission probability in the single-particle
approximation~\cite{f1}. This approach successfully accounts for
unique quantum effects in mesoscopic systems~\cite{FG99}.
The electron wavefunction is assumed to extend coherently across the device
and the two-terminal, linear-response conductance at zero temperature reads

\begin{equation}
\label{meq2}
G(E_{\rm F}) = 2 (e^2/h) T(E_{\rm F}).
\end{equation}

The factor $2$ accounts for spin degeneracy, and $T(E_{\rm F})$ is the
energy-dependent
total transmittance for injected electrons with Fermi energy $E_{\rm F}$.
The transmission function is given by
$T(E)=\sum_{\nu\nu^\prime}|S_{\nu\nu^\prime}(E)|^2$, where $\nu,\nu^\prime$ are
quantum numbers labelling open channels for transport which belong to mutually
exclusive leads. In our case these are the two semi-infinite perfect nanotubes. The attached
molecular system defines a scattering problem, and $S$ is the 
corresponding quantum-mechanical scattering matrix.

We determine the central quantity, $T(E)$, numerically by employing an
efficient algorithm for calculating the Green function for arbitrary
tight-binding Hamiltonians. The $S$-matrix is computed from the Green function
via the Fisher-Lee relation~\cite{FG99}.
This general scattering technique has been recently formulated for
studies of the giant magnetoresistance and is described in detail in
Ref.~\cite{SLJB99}.
It was later applied to a number of problems including the electric conductance in multi-wall
carbon nanotubes~\cite{SKTL00}. The computational scheme proceeds at three
stages: a) the calculation of an effective~(renormalised) interaction between the
electrodes by projecting out the degrees of freedom of the scatterer,
b) the computation of the Green function in the absence of the scatterer,
and c) the use of the Dyson equation to express the required Green function of the
composite system (leads plus scatterer).

\begin{figure}
\begin{center}
\subfigure{\epsfig{file=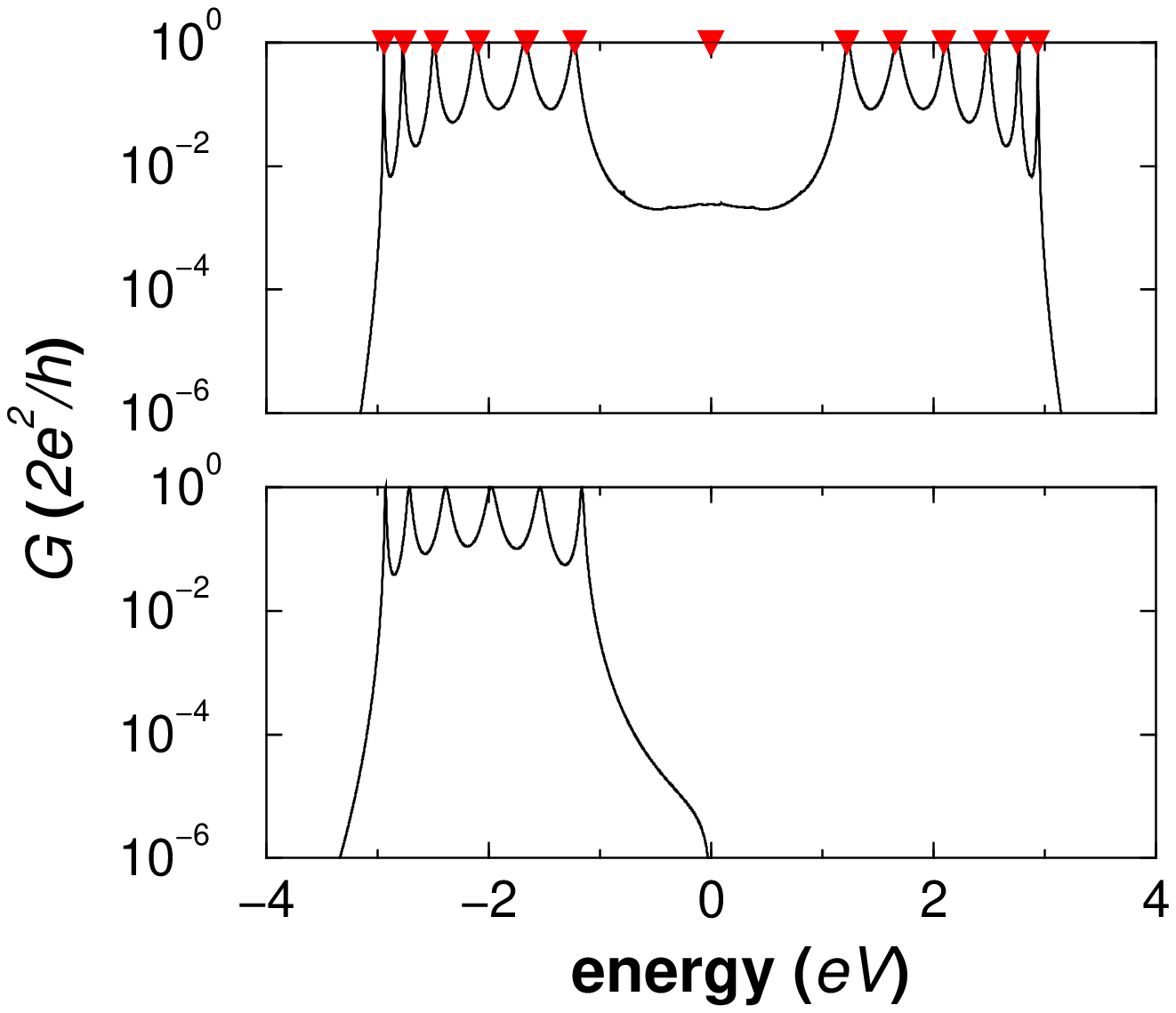, width=.49\linewidth}}
\subfigure{\epsfig{file=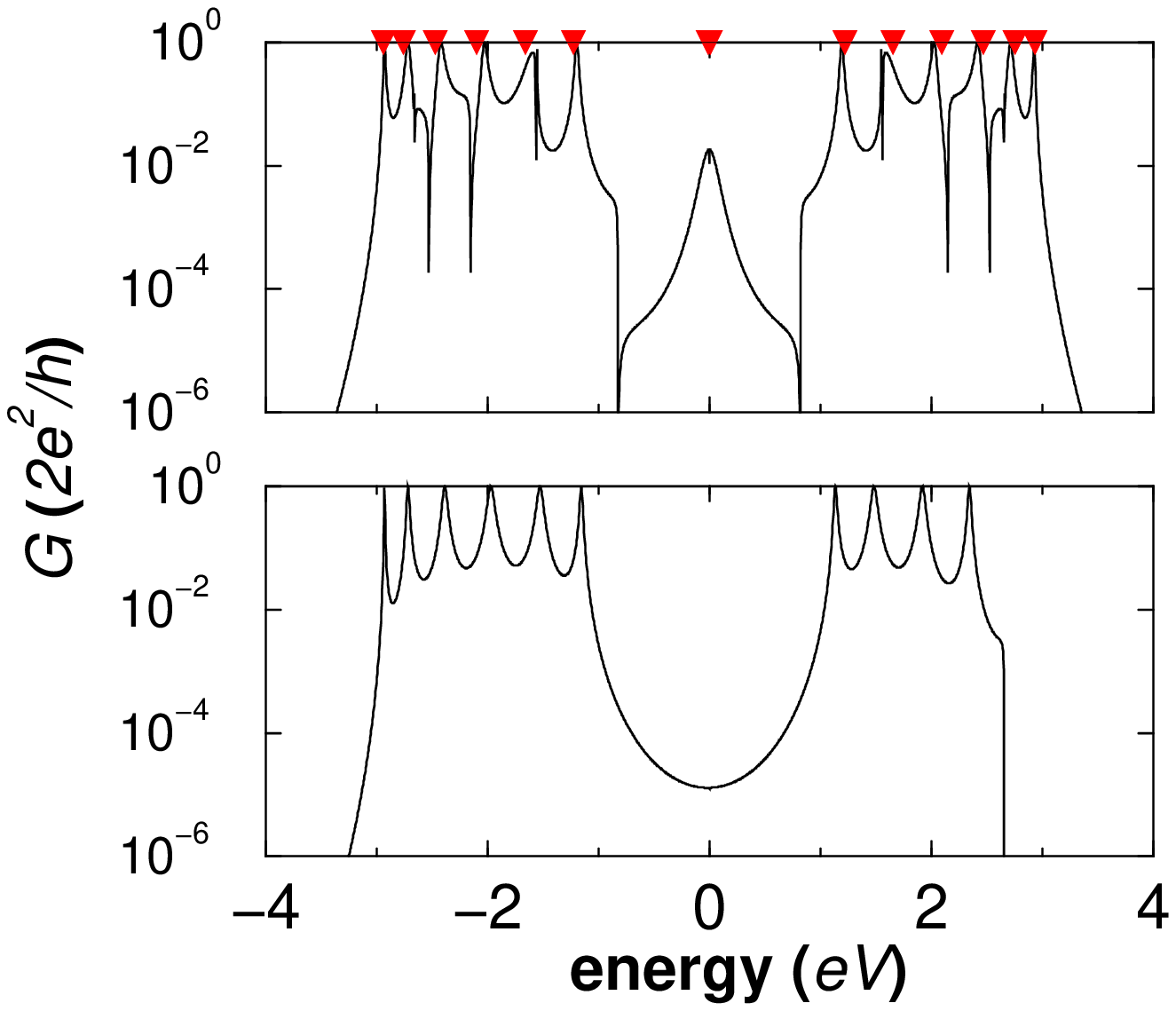, width=.49\linewidth}}
\end{center}

\caption{\label{fig1}
	Typical conductance spectra for an $N\!=\!14$ dimerised chain attached
	to square-lattice tube (left) and (10,10) carbon nanotube
	electrodes (right). An effective multiple contact ($M_c\!=\!20$, bottom)
	acts as a filter selecting a single transport channel in contrast
	to a single-atom contact (top). Symbols
	indicate eigenenergies of an $N\!=\!14$
        iso\-la\-ted dimerised chain. The carbon nanotube and
	square-lattice tube bandwidths are 16eV and 8eV, respectively.
	}
\end{figure}

Typical conductance spectra for the two extreme examples of a single
interfacial atomic contact and multiple
contacts are shown in the left and right panel of Fig.~\ref{fig1}, respectively.
One observes the general resonant character of transport typical of
mesoscopic systems. In particular, the conductance shows resonances of quantum unit~($2 e^2/h$)
height at eigenenergies close to that of the energy spectrum of the isolated
molecular chain which is indicated by symbols~\cite{f4}.
They arise because of back reflections at the molecular wire-electrode interface.
In the SC-scenario all open channels contribute to the transmission, i.e.\
$S_{\nu\nu^\prime}(E)$ is non-zero for any $\nu,\nu^\prime$.
For carbon tubule leads~(upper right panel of Fig.~\ref{fig1}) we observe
additional structure in the conductance spectra.
Additional results show that distinctive features such as
antiresonances reflect the local density of states of the carbon nanotubes and
are signatures of van Hove singularities in the carbon tubules band structure~\cite{ACTA}.
Due to the simplicity of our model no antiresonances from the electronic
molecular structure can occur \cite{EK98}.

The MC-configuration exhibits a profoundly different
behaviour. The conductance vanishes for part of the spectrum
for both square lattice tubes and carbon tubules as shown in the lower panels of Fig.~\ref{fig1}.
In addition, the quantum features associated to the carbon nanotube structure
disappear. The complicated conductance spectrum appears now as a series of
resonant peaks. Detailed analysis of the $S$-matrix elements revealed that only
wavefunctions of the tubes without modulation along the
cross-section circumference allow for transport, thereby, yielding the zero conductance
when such channels are not available. Evidently, this leads to asymmetric conductance
spectra in general. We may say that the MC-configuration acts as a channel filter.
The selection is a consequence of a sum rule that determines the transmission
of each open channel and is related to the spectral density~\cite{YR98}.
It may be viewed as the wavefunction overlap
$\langle \Psi_{\rm wire}| H| \Psi_{\rm L,R} \rangle$
(see Eq.~(\ref{meq1}), due to the nature of coupling only the transverse
profile is important) and, hence, from symmetry considerations it is clear
that the channel selection is a generic feature of cylindrical electrodes. It follows
that channel filtering should approximately prevail also for non-cylindrical,
mesoscopic electrodes with lateral confinement. This has been observed in
additional calculations.

\begin{figure}
\begin{center}
\subfigure{\epsfig{file=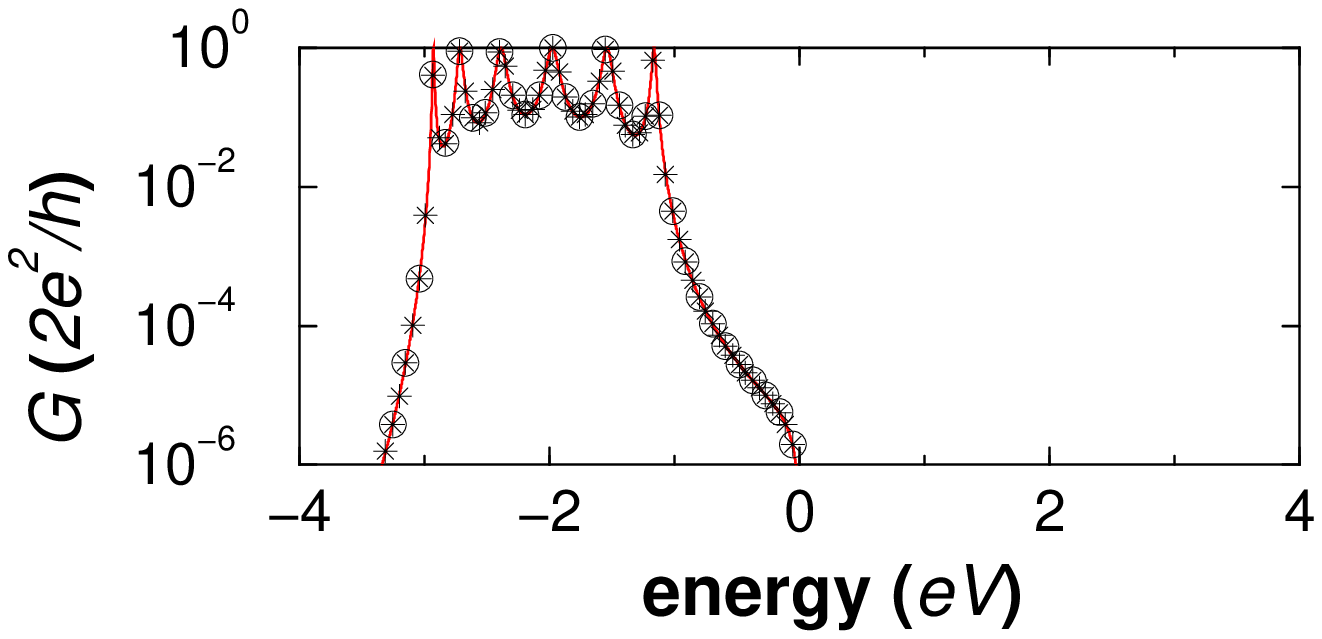, width=.49\linewidth}}
\subfigure{\epsfig{file=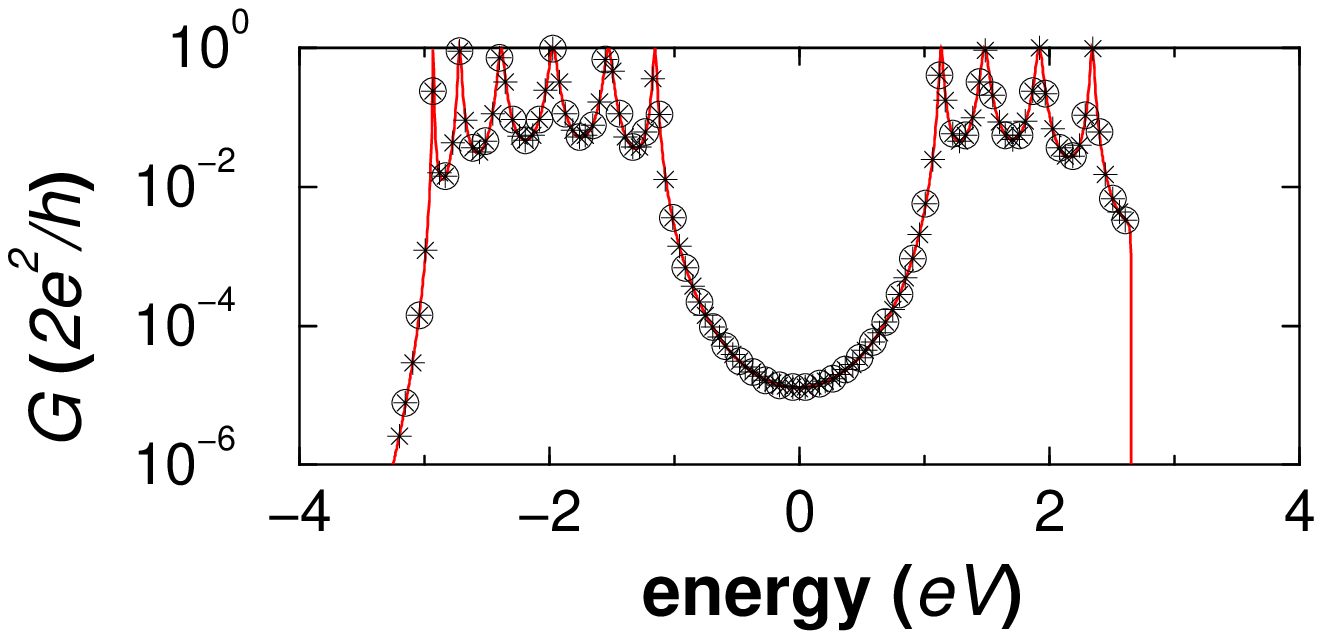, width=.49\linewidth}}
\end{center}

\caption{\label{fig2}
	Superimposed conductance curves with $\Gamma^2 M =$ const
	showing the validity of the sum rule (see text) for the multiple
	contact configuration
	for square lattice tube (left) and carbon nanotube (right) electrodes.
	Symbols and line indicate different $\Gamma$.}
\end{figure}

A further particular feature of the MC configuration is that
the conductance obeys a scaling law. That is, $G = G(\Gamma^2 \cdot D)$,
where  $D$ is the diameter of the tube~(Fig.~\ref{fig2}).
This is a mere contact effect related to the symmetry of the contributing
channel wavefunction that should
hold for any effective coupling with the form considered here and for
any kind of tubule electrode. We can give a qualitative argument by inspection
of the resonant structure of the
conductance spectrum. In a linear-linear plot we see a series of Lorentzian
peaks. According to Breit-Wigner such peaks scale as the square of the coupling
strength times the local density of states. The effective coupling scales as
$\Gamma$ times $M$. The latter relates to the diameter of the tube.
Taking into account the single-channel transport and the fact that the transverse
profile of the contributing channel wavefunction has no nodes and is
normalised, one arrives at a local density of states proportional
to $1/M$ which fixes the scaling law. A more rigorous analysis shows that the conductance depends 
only on $\langle \Psi_{\rm wire}| H| \Psi_{\rm L,R}
\rangle$ via a non-zero molecular wire self-energy induced by the electrodes
~\cite{MKR94a,JCP}. The exact form of the scaling law
readily follows by repeating the above argument.

\begin{figure}[t]
\begin{center}
\subfigure{\epsfig{file=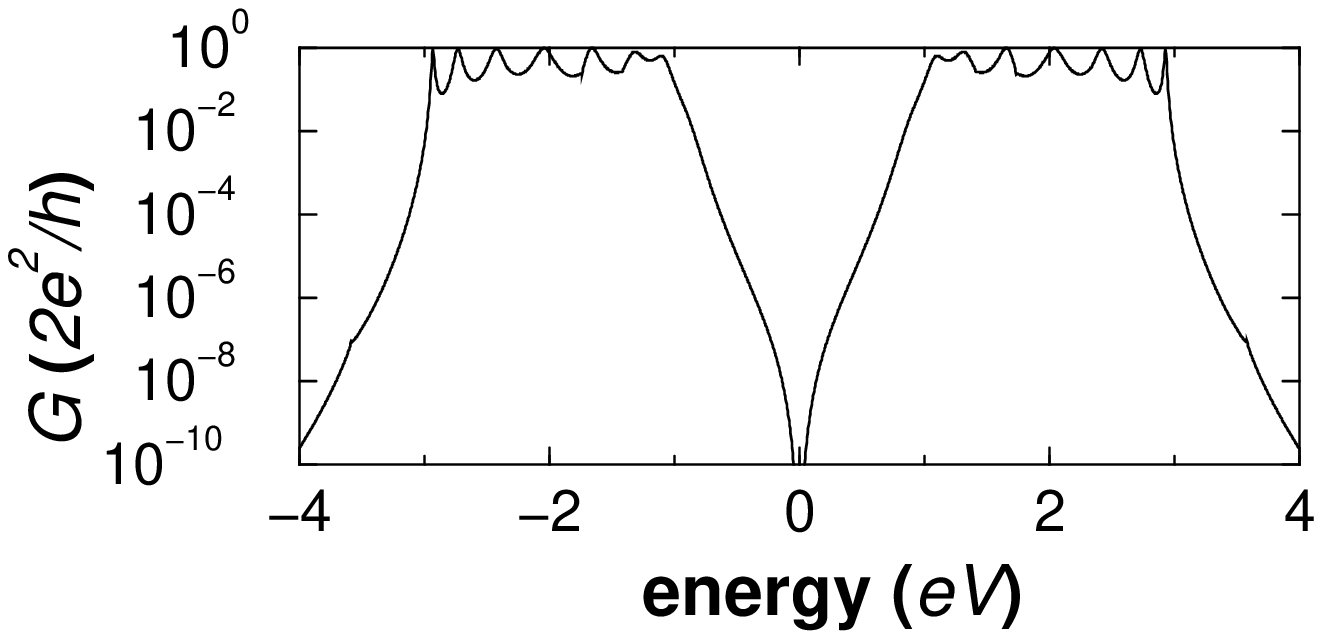, width=.49\linewidth}}
\subfigure{\epsfig{file=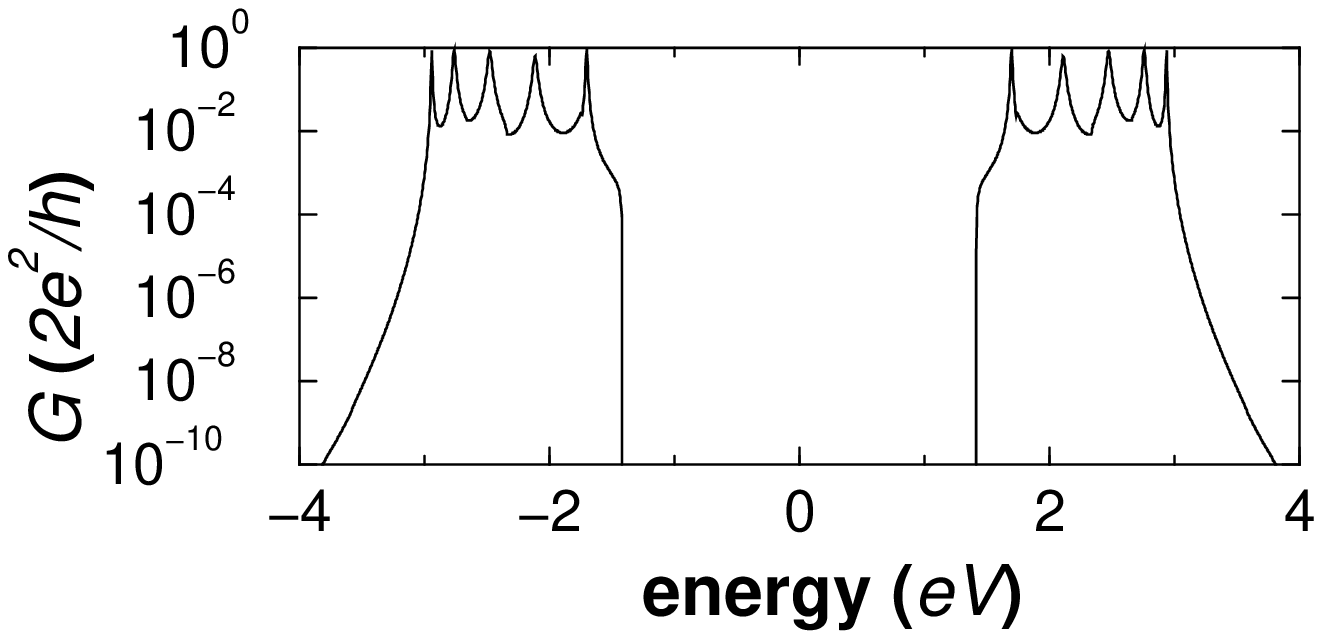, width=.49\linewidth}}
\end{center}

\caption{\label{fig3}
	Conductance spectrum for (9,0) carbon nanotube electrodes with $M_c=1$ (left)
	and $M_c=3$ (right).
	}
\end{figure}

More generally, multiple contacts allow for control of low-dimensional
transport via channel selection with the `intermediate contact' case,
i.e.~$1\!<\!M_c\!<\!M$, exhibiting much richer behaviour. This results from a
combination of the effects we described hitherto.
For completeness we discuss a specific example which supports channel
selection and illustrates once more the importance of the interface in
molecular system-nanoelectrode hybrids. In Fig.~\ref{fig3} typical
conductance spectra for zigzag carbon nanotube electrodes are shown. To interpret these results
we note that transport usually takes place at $E\sim1$eV around the Fermi
energy $E_F=0$. For this part of the spectrum apart from the tunnelling nature
of the conductance owing to the gap of the dimerised chain, we notice a
complete suppression of conductance
for $M_c=3$, due to contact `dimensionality'~\cite{f2}.
The origin of this effect derives from metallic zigzag nanotubes supporting
two degenerate transport channels in this energy region with an antisymmetric
wavefunction within a cell consisting of three atomic sites (along the circumference of the tubule).
Hence, the wire/tube overlap gives a
zero contribution for $M_c=3 \cdot n$ as depicted and non-zero otherwise~\cite{f3}.

In summary, we have studied interfacial effects on the
conductance of a hybrid built by a molecular wire coupled to low-dimensional
leads. We have pointed out immediate consequences of the contact geometry and dimensionality
on electron transport across such systems. These include the possibility of
channel selection and the existence of a scaling law $G=G(\Gamma^2\! \cdot\!
M)$ for devised geometries. Although we presented results for a dimerised
chain such effects are generic. We also demonstrated that a square lattice tube model for
electrodes exhibits the above features. However, replacement by
natural and realistic candidates of molecular electronic circuits,
carbon nanotubes, adds richer structure to the conductance profile. This is
related to the tubule topology which determines the local density of
states. A complementary study of such effects should take into account a
full microscopic description of the molecular system including structural
relaxation and carbon nanotube caps. Another possibility is to consider
multi-wall nanotube leads. In this case, an axial magnetic field of reasonable 
magnitude can be applied which modulates the symmetry of the tube states and,
hence, can act as an external tuning parameter. But perhaps the most
challenging task would be to discuss electron transport for finite temperatures and bias
voltages, phase-breaking mechanisms, and charging effects and see how the
above features are modified.

We would like to thank the organisers of the `Molecular Electronics 2000' conference
for giving us the opportunity to present this work. We also thank
A. Bachtold, C.J. Lambert, and S. Sanvito for comments and discussions.
G.F. acknowledges support by P. Fulde for participation at this conference.


\newpage

\newpage

\newpage

\newpage

\ecols

\begin{thebibliography}{10}
\bibitem{f4}
With increased resolution one sees that there are two eigenenergies
lying symmetrically to zero energy which are present only in the case of open
boundary conditions. They correspond to states with identical probability densities peaked at
the chain end-atoms and suppressed elsewhere explaining the tunnelling nature
of the conductance.
\bibitem{f2}
The position of the contacts does not influence conductance for zigzag carbon 
tubule leads, while for armchair carbon tubules there is no distinction only
if $M_ c$ is odd. This is explained by the structure of the carbon nanotube
unit cell.
\bibitem{f3}
For armchair carbon tubules,
conductance is always finite but for $M_c$ even,
only one channel instead of two is contributing. Arguments similar to those
given for zigzag nanotubes hold.
\end{thebibliography}
\end{document}